\documentclass[%
 aip,
 amsmath,amssymb,
 reprint,%
]{revtex4-1}

\usepackage{graphicx}
\usepackage{dcolumn}
\usepackage{bm}

\usepackage[utf8]{inputenc}
\usepackage[T1]{fontenc}
\usepackage{mathptmx}
\usepackage{etoolbox}

\makeatletter
\def\@email#1#2{%
 \endgroup
 \patchcmd{\titleblock@produce}
  {\frontmatter@RRAPformat}
  {\frontmatter@RRAPformat{\produce@RRAP{*#1\href{mailto:#2}{#2}}}\frontmatter@RRAPformat}
  {}{}
}%
\makeatother
\begin{document}

\preprint{AIP/123-QED}

\title{Superconducting aluminum heat switch with 3 n$\Omega$ equivalent resistance}
\author{James Butterworth}
\affiliation{
Air Liquide Advanced Technologies 2, rue de Clemenciere, BP 15, 38360 Sassenage, France
}
\author{S\'{e}bastien Triqueneaux}
\affiliation{
Univ. Grenoble Alpes, CNRS, Grenoble INP, Institut Neel, 38000 Grenoble, France
}
\author{\v{S}imon Midlik}
\affiliation{
Charles University, Ke Karlovu 3, 121 16, Prague, Czech Republic
}
\author{Ilya Golokolenov}
\author{Anne Gerardin}
\author{Thibaut Gandit}
\author{Guillaume Donnier-Valentin}
\author{Johannes Goupy}
\author{M. Keith Phuthi}
\affiliation{
Univ. Grenoble Alpes, CNRS, Grenoble INP, Institut Neel, 38000 Grenoble, France
}
\author{David Schmoranzer}
\affiliation{
Charles University, Ke Karlovu 3, 121 16, Prague, Czech Republic
}
\author{Eddy Collin}
\author{Andrew Fefferman}
\email{andrew.fefferman@neel.cnrs.fr}
\affiliation{
Univ. Grenoble Alpes, CNRS, Grenoble INP, Institut Neel, 38000 Grenoble, France
}

\date{\today}

\begin{abstract}
Superconducting heat switches with extremely low normal state resistances are needed for constructing continuous nuclear demagnetization refrigerators with high cooling power. Aluminum is a suitable superconductor for the heat switch because of its high Debye temperature and its commercial availability in high purity. We have constructed a high quality Al heat switch whose design is significantly different than that of previous heat switches. In order to join the Al to Cu with low contact resistance, we plasma etched the Al to remove its oxide layer then immediately deposited Au without breaking the vacuum of the e-beam evaporator. In the normal state of the heat switch, we measured a thermal conductance of $8 T$ W/K$^2$ which is equivalent to an electrical resistance of 3 n$\Omega$ according to the Wiedemann-Franz law. In the superconducting state we measured a thermal conductance that is $2\times10^6$ times lower than that of the normal state at 50 mK.
\end{abstract}

\maketitle

\section{\label{sec:Intro}Introduction}

Cryogen-free (``dry'') dilution refrigerators have become popular in recent years because of their large experimental space, their automated operation and the cost of helium. The advantages of dry dilution fridges have allowed a large community to access the range $T\gtrsim10$ mK. However, several fields of research require still lower temperatures. \cite{Albash17,Yurttagul19,Lotnyk21,Nguyen21} For example, recent work demonstrated that the energy relaxation time of fluxonium qubits increases as their transition frequency is decreased at least down to 200 MHz.\cite{Nguyen19} It is desirable to cool such qubits well below 10 mK so as to minimize the excited state populations. Sub-mK temperatures in condensed matter systems are achieved using adiabatic nuclear demagnetization refrigerators (NDR), which usually include liquid helium baths and have a relatively small experimental space. A few laboratories have realized cryogen-free, single shot NDR, reaching impressive minimum temperatures below 100 $\mu$K.\cite{Batey13,Todoshchenko14,Palma17,Yan21} However, these dry NDR had relatively high heat leaks of a few nW. This results in a limited autonomy of a few days below 1 mK as well as pre-cooling times that are at least as long,\cite{Yan21} making these systems impractical for many applications.

The recently proposed continuous nuclear demagnetization refrigerators (CNDR), which are based on multiple nuclear demagnetization stages arranged in a series or parallel configuration, would allow ultra-low temperature researchers to take full advantage of cryogen-free technology.\cite{Toda18, Schmoranzer19} According to numerical calculations, it is possible to construct a CNDR with a cooling power of tens of nW at 1 mK.\cite{Schmoranzer20} However, building a CNDR is challenging in part because the two nuclear stages must be linked by a very low thermal resistance with equivalent resistance in the n$\Omega$ range. This thermal resistance limits the rate at which the CNDR can be cycled and consequently limits its cooling power. The simpler series configuration of the CNDR is particularly sensitive to the resistance of the thermal link.\cite{Schmoranzer20}

An important part of a CNDR thermal link is the heat switch. Only superconducting heat switches simultaneously meet the requirements of low heat dissipation and high switching ratio. This type of heat switch relies on the fact that, well below the critical temperature $T_{c}$ of a superconductor, only phonons contribute significantly to its thermal conductance. This is because the density of electronic quasiparticle excitations, which transport heat, decreases exponentially as the temperature is lowered below $T_{c}$.\cite{Bardeen59} Thus for $T\ll T_{c}$ the thermal conductance of superconductors can be \textquotedblleft switched\textquotedblright\ by applying a magnetic field that is large enough to destroy the superconductivity. The switching ratio, i.e. the ratio of the normal state thermal conductance to that of the superconducting state, can exceed 10$^{6}$.\cite{Mueller78}

Superconducting heat switches have been made of Al, In, Pb, Sn and Zn. The designs and normal state conductances of several heat switches using different superconductors are tabulated in Ref. \onlinecite{Triqueneaux21}. Aluminum is an attractive material for constructing heat switches because of its relatively high Debye temperature $\Theta _{D}=394$ K, which yields a relatively low phonon conductivity at a given temperature and phonon mean free path. Also, it is commercially available in 6N purity, yielding a high normal state conductance; it has a lower critical magnetic field than In, Pb and Sn; and it is relatively strong. Mueller \textit{et al}. pioneered the use of Al in heat switches, presenting a solution to the problem of Al surface oxide.\cite{Mueller78} In particular, an elaborate chemical procedure was used to gold plate contact areas at the ends of 20 Al foils. These Al foils were then interleaved with two sets of 21 gold-plated Cu foils. The Weidemann-Franz equivalent electrical resistance of their heat switch was 15 n$\Omega $ in the normal state. To our knowledge, this heat switch had the lowest normal state thermal resistance until now.\cite{Triqueneaux21}

In this work, we report thermal measurements of an aluminum heat switch with a much lower normal state thermal resistance equivalent to 3 n$\Omega $. The chemical gold plating procedure of Ref. \onlinecite{Mueller78} has been replaced with plasma etching of the aluminum followed by gold deposition without breaking vacuum.\cite{Triqueneaux21} Furthermore, an adequate superconducting state thermal resistance for use in a CNDR is achieved using a single piece of aluminum, as opposed to the several foils employed in Ref. \onlinecite{Mueller78}. This technology will be instrumental to the construction of a CNDR with high cooling power at sub-mK temperatures.

\section{\label{sec:setup}Experimental Setup}
\begin{figure}
\includegraphics[width=\linewidth]{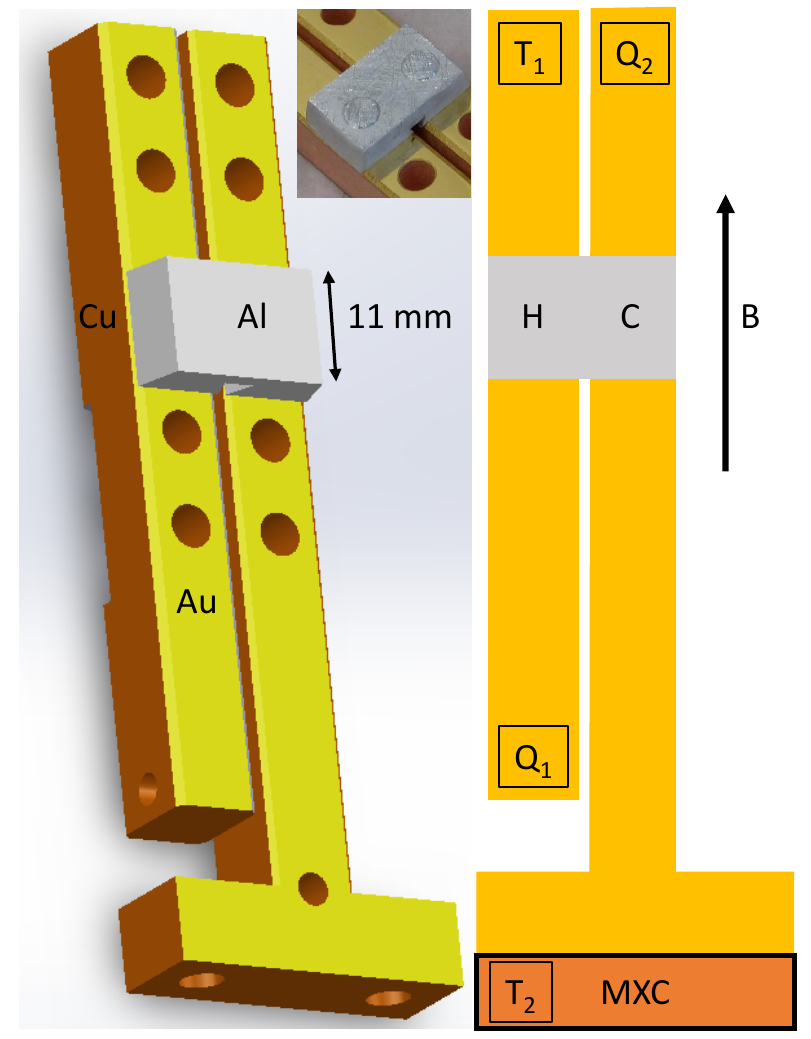}
\caption{\label{fig:setup} Left: The heat switch consists of two Au-plated Cu legs joined to a Au-plated Al piece. Inset: Photo of the Al element and part of the Au-plated Cu legs. Right: The positions of the heaters (Q$_1$, Q$_2$) and thermometers (T$_1$, T$_2$) and the direction of the magnetic field $B$ are indicated. The hot and cold sides of the heat switch are labelled ``H'' and ``C'', respectively, and the mixing chamber plate is labelled ``MXC''.}
\end{figure}

Our heat switch is made of a single Al piece joined to two Cu legs that formed the ``hot'' and ``cold'' sides of the heat switch (Fig. \ref{fig:setup}). The joint is a gold-plated pressed contact as explained in detail in Ref. \onlinecite{Triqueneaux21}. The Al portion was cut from a 99.9999\% pure rod\footnote{AlfaAesar} by spark erosion. The slot in the Al in the region between the two Cu legs forms a constriction that moderately limits the thermal conductance in the superconducting state. We estimate that the effective section/length of the Al element is 5 mm. Holes were drilled into the Cu legs so that we could attach thermometers and heaters as shown in Fig. \ref{fig:setup}. The thermometers were Speer carbon resistors calibrated using a $^3$He melting curve thermometer. We used 120 $\Omega$ strain gauges as heaters because they have a nearly temperature independent electrical resistance and are easy to obtain.\footnote{RS Pro} The holes drilled in the base of the T-shaped Cu piece were used to attach the heat switch to the Au-plated mixing chamber plate of a dilution refrigerator. The surface of that Cu piece in contact with the MXC plate was coated with Au using Dalic brush plating solution. The thermal resistance of this contact was not optimized since, for the purposes of our test, it only had to be (1) low enough so that the cold side of the heat switch remained close to the temperature of the mixing chamber even at high applied powers and (2) small compared with the thermal resistance of the Al in its superconducting state (see below). A superconducting coil surrounded the Al part of the heat switch and was used to apply a magnetic field $B=10.6$ mT that destroyed the superconductivity of the Al.

\section{\label{sec:results}Results}

\begin{figure}
\includegraphics[width=\linewidth]{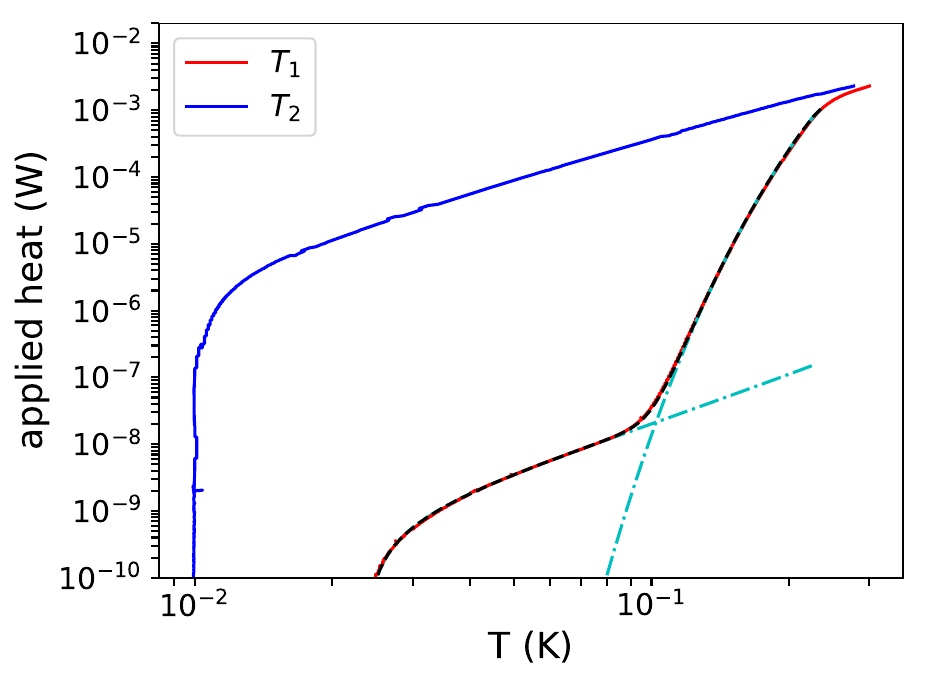}
\caption{\label{fig:Qdotsc} The temperatures of thermometers T$_1$ and T$_2$ were measured for different amounts of heat dissipated in heater Q$_1$ with the heat switch in its superconducting state. The two blue dash-dot curves correspond to the two fitting functions (see text) and the black dashed curve is their sum.}
\end{figure}

At zero applied magnetic field, so that the Al was in its superconducting state, we measured the temperatures $T_1$ and $T_2$ of thermometers T$_1$ and T$_2$, respectively, for a range of powers $\dot{Q}_{app}$ dissipated in heater Q$_1$ (Fig. \ref{fig:Qdotsc}). The thermal resistance was almost entirely in the Al, so that $T_1$ ($T_2$) closely approximates the temperature at the hot (cold) ends of our Al element, except at the highest applied powers (see below). We fitted the empirical function $\dot{Q}_{app}=aT_1^3+bT_1^2-p$ to the measurements $\left(\dot{Q}_{app},T_1\right)$ for $T_1<100$ mK and obtained best fit values $a=7.1~\mu$W/K$^3$, $b=1.4~\mu$W/K$^2$ and $p=0.89$ nW. Here $p$ is a temperature independent parasitic heat load. For $T_1>100$ mK we used the fitting function \cite{Pekola21} $\dot{Q}_{app}=\int_{0}^{T_{1}}2\Delta^2/(e^2RT)\exp\left[-\Delta/\left(k_BT\right)\right] dT$ where $e=1.6\times10^{-19} C$ is the fundamental charge, $\Delta=\alpha k_B T_c$ and $T_c=1.2$ K. Upon optimizing the fit we obtained $R=2$ n$\Omega$ and $\alpha=1.53$. In Fig. \ref{fig:Qdotsc} these two fitting functions are shown as dash-dot curves and their sum is shown as a dashed curve.

\begin{figure}
\includegraphics[width=\linewidth]{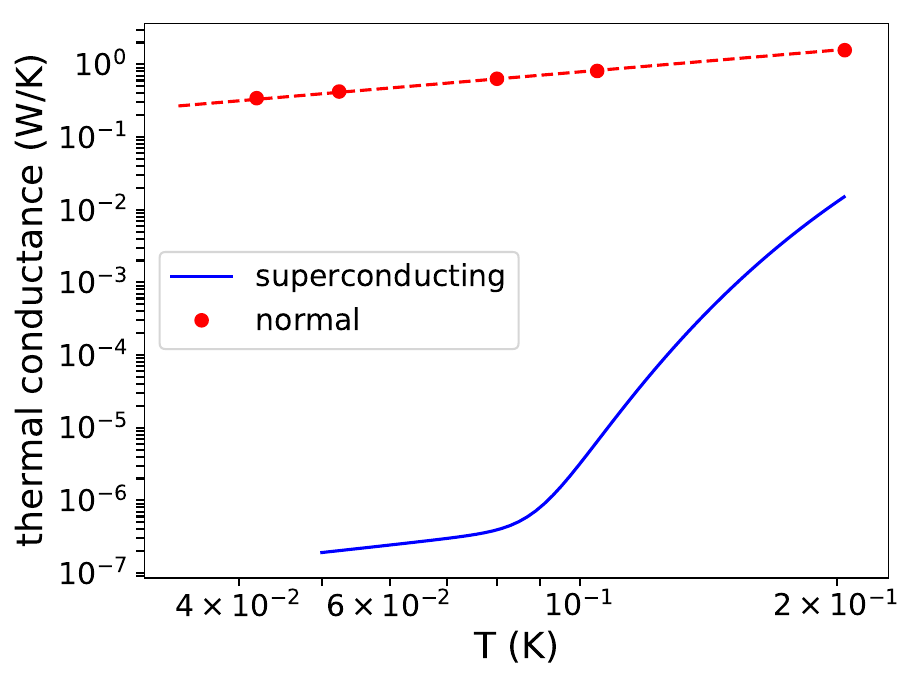}
\caption{\label{fig:k} Thermal conductances of the heat switch: The conductance in the superconducting state is derived from Fig. \ref{fig:Qdotsc} (see text), the points are measurements of the normal state conductance, and the dashed line is a linear fit corresponding to a Wiedemann-Franz equivalent resistance of 3 n$\Omega$.}
\end{figure}

The corresponding thermal conductance was obtained from the temperature derivative of the total fitting function in the range 50 mK$<T_1<$200 mK (Fig. \ref{fig:k}). In this range of $T_1$, $T_2$ is sufficiently small so that it does not significantly limit the heat flow through the heat switch. Furthermore, the measured thermal conductance in the superconducting state of the heat switch was much less than that of the MXC-heat switch joint up to 200 mK. In particular, the latter was equivalent to 65 n$\Omega$. Therefore the measured thermal resistance is almost entirely due to the Al element in our chosen $T_1$ range. We verified that MXC joint resistance was negligible by calculating the temperature $T_i$ between the Al element and the joint with the MXC plate at each applied heat level. We then recalculated the superconducting state thermal conductivity integral for $T_1>100$ mK by changing the lower limit from zero to $T_i$. Increasing the lower limit of the thermal conductivity integral in the superconducting state from zero, as assumed above, to $T_i$ resulted in a negligible change in the thermal conductivity integral.

With the Al in the normal state, we measured the thermal resistance $(T_H-T_C)/\dot{Q}$, where $T_H$ and $T_C$ are the temperatures of the Cu near the Al joint on the hot and cold sides of the switch, respectively, with power $\dot{Q}$ flowing through the heat switch. This thermal resistance includes the contributions of the two Cu/Au/Al joints and the interior of the aluminum. We made this measurement using the two heater technique (Ref. \onlinecite{Dhuley19} and references therein). The thermometer, T$_1$, was located on an extension of the hot Cu leg through which almost none of the applied heat flowed (Fig. \ref{fig:setup}). Consequently thermometer T$_1$ was nearly at temperature $T_H$ when power $\dot{Q}$ was dissipated in heater Q$_1$. When heater Q$_1$ was switched off and power $\dot{Q}$ was dissipated in heater Q$_2$, T$_1$ was nearly at temperature $T_C$ since very little heat was flowing through the Cu on the hot side or through the Al. In order to ensure that the same power $\dot{Q}$ was dissipated in each of the heaters we adjusted the current applied to each heater to obtain the same temperature reading on thermometer T$_2$. The two heater technique is accurate only if the cooling power of the mixing chamber does not drift significantly when one switches from one heater to the other. We therefore verified the accuracy of our measurement of the thermal resistance by repeating it several times, and we found that it was reproducible. The corresponding thermal conductance $k_n=\dot{Q}/(T_H-T_C)$ is plotted in Fig. \ref{fig:k}. Electron-phonon scattering is negligible compared with temperature independent scattering mechanisms at these temperatures, yielding a temperature independent electrical resistance $R$. As expected from the Wiedemann-Franz law, $k_e=L T/R$, where $k_e$ is the electronic thermal conductance and $L=2.44\times10^{-8}$ W $\Omega$/K$^2$ is the Lorenz number, the measured normal state conductance is linear in temperature. In particular, the Wiedemann-Franz equivalent resistance is $R=3$ n$\Omega$.

We note that the 2 n$\Omega$ that optimized the fit to the superconducting state conducted heat (Fig. \ref{fig:Qdotsc}) corresponds to the normal state resistance of the interior of the Al. Thus the combined Wiedemann-Franz equivalent resistance of our two joints is approximately 1 n$\Omega$. Electrical measurements of similar joints at 4 kelvin yielded comparable resistances.\cite{Triqueneaux21}

In order to quantify the eddy current heating due to the applied magnetic field, we ramped the field in a triangle pattern at a rate of 0.03 mT/sec while maintaining the heat switch in the superconducting state. This ramp rate is adequate for controlling the heat switch state of a CNDR, even if it is necessary to ramp the field from zero to the 10 mT critical field of aluminum, since the CNDR cycle is expected to last more than one hour.\cite{Schmoranzer20} Based on the resulting value of the temperature $T_1$ and our measurement of the superconducting state heat conduction (Fig. \ref{fig:Qdotsc}), we inferred 1 nW of eddy current heating. This is comparable to the parasitic heat loads of dry NDR cited in the introduction and less than the expected cooling power of the CNDR.\cite{Schmoranzer20}

\section{Conclusion}
The heat switch described here fulfills the requirements for use in a CNDR. In particular, its normal state thermal conductance is more than sufficient. While the contact resistance between the heat switch and the mixing chamber plate was much greater than that of the Al/Au/Cu joint in our test setup, copper to copper joints with contact resistances of less than 10 nOhms can be readily achieved: The important factors are the compression force\cite{Blondelle14} and the use of a gold or indium interfacial layer.\cite{Okamoto90} Furthermore, the superconducting state thermal conductance of the heat switch is sufficiently small so that the parasitic heat loads of a few nW typically observed in cryogen-free demagnetization refrigerators will dominate leakage through the heat switch from the mixing chamber at $\approx$10 mK to the nuclear stage at $\approx$1 mK. If necessary, further reduction of the superconducting state thermal conductance could be achieved by adjusting the aspect ratio of the Al constriction. Furthermore, the geometry of the heat switch is such that eddy current heating is sufficiently small. The high normal state conductance of our heat switch will facilitate the construction of a powerful CNDR compatible with a cryogen-free dilution refrigerator, thereby extending the benefits of dry fridges to the microkelvin temperature range.

\section{Acknowledgements}
We acknowledge support from the ERC StG grant UNIGLASS No.714692 and ERC CoG grant ULT-NEMS No. 647917. The research leading to these results has received funding from the European Union's Horizon 2020 Research and Innovation Programme, under Grant Agreement no 824109. Deposition of the Au film took place at the Plateforme Technologique Amont (PTA) of Grenoble. MKP's international research experience with MIT was made possible by the GIANT International Internship Programme (GIIP), GIANT Innovation Campus, Grenoble, France.

\section{Data Availability}
The data used here is available at Ref. \onlinecite{data}.

\end{document}